\documentclass[aps,prl,twocolumn,amsfonts,amssymb,amsmath,showpacs,
floatfix,nofootinbib,groupedaddress,superscriptaddress,citesort]{revtex4}
\usepackage{mathrsfs}
\usepackage{amsfonts}
\usepackage{amstext}
\usepackage{amsmath}
\usepackage{amssymb}
\usepackage{float}
\usepackage[usenames]{xcolor}
\usepackage[dvips]{graphicx}
\def\qed{\leavevmode\unskip\penalty9999 \hbox{}\nobreak\hfill
     \quad\hbox{\leavevmode  \hbox to.77778em{%
              \hfil\vrule   \vbox to.675em%
               {\hrule width.6em\vfil\hrule}\vrule\hfil}}
     \par\vskip3pt}

\def\ra{\rangle}
\def\la{\langle}

\def\no{\nonumber}

\begin{document}
\title{Quantum correlations with vacuum ambiguity in de Sitter space}

\author{Jun Feng}
\affiliation{Beijing National Laboratory for Condensed Matter Physics,
Institute of Physics, Chinese Academy of Sciences, Beijing 100190, P. R. China}
\author{Cheng-Yi Sun}
\affiliation{Institute of Modern Physics, Northwest University, Xian 710069, P. R. China}
\author{Wen-Li Yang}
\affiliation{Institute of Modern Physics, Northwest University, Xian 710069, P. R. China}
\author{Yao-Zhong Zhang}
\affiliation{School of Mathematics and Physics, The University of Queensland, Brisbane, Qld 4072, Australia}
\author{Heng Fan}
\email{hfan@iphy.ac.cn}
\affiliation{Beijing National Laboratory for Condensed Matter Physics,
Institute of Physics, Chinese Academy of Sciences, Beijing 100190, P. R. China}

\begin{abstract}
We study the quantum correlations of free scalar field with vacuum ambiguity of de Sitter space.
We show the occurrence of degradation of quantum entanglement and quantum discord between field modes
for inertial observer in curved space due to the radiation associated with cosmological horizon.
In particular,
we find that quantum correlations can be used to encode infinite de Sitter invariant vacua,
which correspond to infinite set of possible physical worlds.
This may provide a superselection rule of physical vacuum via quantum information tasks.
We also discuss the simulation of such quantum effects of vacuum ambiguity in ion trap experiments.

\end{abstract}
\pacs{03.67.Mn, 03.65.Ud, 04.62.+v,}
\maketitle

\def\bea{\begin{eqnarray}}
\def\eea{\end{eqnarray}}
\def\be{\begin{equation}}
\def\ee{\end{equation}}

\def\no{\nonumber}

\emph{Introduction.}---Quantum information processing, which is based on principles of quantum mechanics, promises algorithms which surpass their
classical counterparts and unconditional secure for quantum communication.
Relativity is another cornerstone of modern physics, and the role of relativity in quantum information
processing has also attracted much attention,
see Ref. \cite{Peres} for a review. Besides its importance in such as satellite-based global
quantum communication \cite{satellite} and optical clocks \cite{clock},
it is believed that investigation of quantum information in a relativistic framework can also shed
light on the study of information paradox of black hole \cite{BH1} and cosmological evolution \cite{cos}.
Moreover, it is possible to detect such relativistic
quantum effects in laboratory systems like ion traps \cite{ion} and cold-atom Bose-Einstein condensation \cite{BEC},
which makes relativistic quantum information (RQI) processing possible for real applications.

A novel phenomenon in RQI is that quantum correlations of physical system mostly entanglement are highly observer-dependent \cite{obdepen}. For a bipartite entangled system in flat space, it was shown \cite{flat,discord3} that an accelerated observer would experience a decrement of the quantum correlations he shared initially
with an inertial partner. This is because that the uniformly accelerated observer has no access to information beyond casual horizon appearing which separates all events in flat space to disconnected Rindler regions. Therefore the information-lost results in the so-called Unruh effect which claims the detection of a thermal bath for accelerated detector in flat space \cite{unruh}.
Such decoherence phenomena should exist for any other kinds of \emph{causal horizons}
related with thermal radiation, for example, in the case of \emph{event horizon} of black hole.
The degradation of quantum correlations
were provoked by the Hawking effect \cite{BH3} and quantum information can escape from a black hole by a process akin to teleportation \cite{BH1} but without the classical information transmitted.

As an idealization of inflation regime in cosmology, de Sitter space contains another important causal horizon, i.e. the \emph{cosmological event horizon}. It was shown \cite{GH} that even an inertial observer in de Sitter space should detect a thermal bath emanating from the de Sitter event horizon with temperature $T=H/2\pi$, where $H$ is Hubble scale. This so-called Gibbons-Hawking effect would also result a lost of entanglement and can be distinguished with a thermal spectrum in Minkowski space by entangling power with respect to local inertial detectors \cite{cos2}.
It is also proposed \cite{cos1}
from the viewpoint similar as the quantum information methodology
a understanding on the appearance of classicality of quantum fluctuations after
inflationary epoch which eventually seeds the large-scale structure of present universe.

In this Letter, we will analyze the quantum correlations of free quantum field from the vacuum ambiguity in de Sitter space,
 which is not yet studied. By ambiguity, it means there exists an infinite family of vacuum states which are de Sitter invariant (i.e. invariant under spacetime isometries) and consistent with CPT invariance \cite{MA1}. Since all these so-called $\alpha$-vacua can be chosen as initial state of early universe (e.g. the Bunch-Davies vacuum \cite{BDvacuum} or Euclidean vacuum corresponding Re$\alpha=-\infty$), one in fact obtains an infinite set of possible physical worlds.
 Those all possible physical worlds should be specified by the signature of trans-Planckian physics from observation, such as the primeval anisotropies in the cosmic microwave background radiation (CMBR) of early universe performed in COBE satellite experiment \cite{COBE,transplanck}. We will show that this vacuum ambiguity has a significance in quantum information tasks by means that it has a directly consequence on
 the behaviors of quantum correlations.

\emph{Vacuum ambiguity in de Sitter space.}---We start from the vacuum definition for de Sitter space which can be embedded in 5-dimensional Minkowski space as the hyperboloid, $-z_0^2+z_1^2+z_2^2+z_3^2=H^{-2}\equiv l^2$, where $l$ is curvature radius. By construction, it admit a isometry group $SO(1,4)$ and enjoys the same degree of symmetry as Minkowski space. A quantized free scalar field in de Sitter space has a mode expansion
\be
\phi(x)=\sum_k[a_ku_k(x)+a_{-k}^\dag u_{-k}^*(x)]
\label{expansion}
\ee
The corresponding vacuum state is defined by $a_k|vac\rangle=0$ and should respects the the spacetime isometry. To specifying mode functions $\{u_k\}$, one must solve the Klein-Gorden equation with some affiliated coordinate systems for different observers \cite{BD}.

\begin{figure}[hbtp]
\includegraphics[width=.35\textwidth]{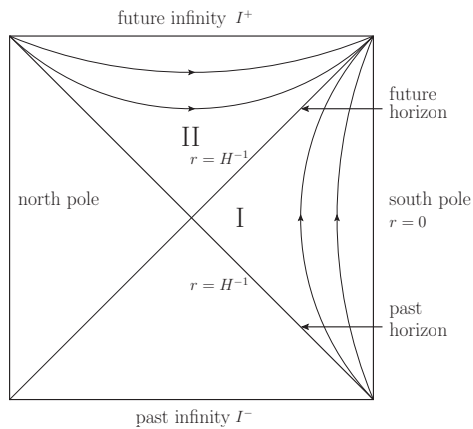}
\caption{The Penrose-Carter diagram of de Sitter space \cite{GH}. The arrowed curves are orbits of the Killing vector. The static observer situated at $r=0$ would observe the event horizon at $r=H^{-1}$.}
\label{penrose}
\end{figure}

In inflation regime, one always adopts planar coordinates which reduce the de Sitter metric as
\be
ds^2=-dt^2+e^{2tH}d\vec{x}^2=\frac{1}{(H\xi)^2}(-d\xi^2+d\rho^2+\rho^2d\Omega^2)
\label{planar}
\ee
where $\xi=-e^{-Ht}/H$ is conformal time. The coordinates cover the upper right triangle of the Carter-Penrose diagram (both region I and II), as depicted in Fig.\ref{penrose}.
The mode functions in this coordinates satisfy the boundary condition, $\partial_\xi u_k(\xi,\vec{x})=-iku_k(\xi,\vec{x})$, in the $\xi\rightarrow-\infty$ limit. This means these Euclidean modes reduce to  the adiabatic modes defined for flat space in the limit of infinite curvature radius. The corresponding Euclidean vacuum state $|0^{(E)}_k\ra$ specified by $a_k^{(E)}|0^{(E)}_k\ra=0$ should also matches the conformal vacuum of Minkowski space in same limit. Since its invariance under de Sitter isometries \cite{BDvacuum}, the Euclidean vacuum (or Bunch-Davies vacuum as called in cosmology) is always chose as initial state to estimate the primordial power spectrum of inflationary perturbations.

Consider a new set of mode basis related with Euclidean one by Mottola-Allen transformation \cite{MA1}
\be
u_k^{(\alpha)}(\xi,\vec{x})=N_\alpha [u_k^{(E)}(\xi,\vec{x})+e^{\alpha} u_{-k}^{(E)*}(\xi,\vec{x})]
\ee
where $\alpha$ is arbitrary complex number with Re$\alpha<0$, $N_\alpha=1/\sqrt{1-e^{\alpha+\alpha^*}}$. The new vacuum states are defined as $a_k^{(\alpha)}|0^{(\alpha)}_k\ra=0$, where $a_k^{(\alpha)}=N_\alpha [a_k^{(E)}-e^{\alpha^*} a_{-k}^{(E)\;\dagger}]$ are corresponding annihilation operators.

These so-called $\alpha$-vacua preserve all de Sitter isometries and hence are also \emph{natural} vacuum states for theory. As Re$\alpha\rightarrow-\infty$, $a_k^{(\alpha)}\rightarrow a_k^{(E)}$ which indicates the Euclidean vacuum as an element of $\alpha$-vacua. Since the time-reversal invariance of theory requires $\alpha$ be real, we henceforth adopt $\alpha=\mbox{Re}\alpha$ for simplicity.

To study the thermality of these de Sitter invariant vacua, one introduces static coordinates for inertial observer and the spacetime metric becomes
\be
ds^2=-(1-r^2H^2)dt^2+(1-r^2H^2)^{-1}dr^2+r^2d\Omega^2
\label{static}
\ee
which covers the region I in Fig. \ref{penrose} and possess a event horizon $r=H^{-1}$ for an observer situated at $r=0$. Following the trajectory of the Killing vector $\partial_t$, the natural choice of quantum field modes can be made for comoving observers. Denoting the field modes and associated annihilation operators as $v_k^{(I)}$ and $b_k^{(I)}$, the particular geodesic observer at $r=0$ should define a static vacuum state by $b_k^{(I)}|0^{(I)}\ra=0$ for all $k$. It should be noted that to construct a complete basis of field expansion (\ref{expansion}) one needs to combine another set of modes $v_k^{(II)}$ defined in region II. However, the event horizon makes an static observer in region I have no access to field modes in region II and must be traced over.

Since the field modes in two coordinates systems do not coincident, the associated annihilator operators can be related with each other by Bogoliubov transformations, $a^{(i)}_k=\cosh r b_k^{(I)}-\sinh r b_{-k}^{(II)\;\dag}$,
where the superscript $i$ refers Euclidean $E$ or general $\alpha$. With squeezing operator $S(r)=\exp[r(b_k^{(I)\dag} b_{-k}^{(II)\dag}-b^{(II)}_kb_{-k}^{(I)})]$, the vacuum for comoving observer in conformal time $\xi$ can be realized as squeezed states
\be
|0^{(i)}_k\ra=\mbox{sech}\;r\sum_{n=0}^\infty\tanh^nr|n_k^{(I)};n_{-k}^{(II)}\ra
\label{2mode}
\ee
which means particles created in pairs one on either side of event horizon, only one in Region I should be detected as de Sitter radiation by inertial observer at $r=0$. For Euclidean vacuum $|0^{(E)}_k\ra$, the well-known Gibbons-Hawking effect gives that, $\tanh^2r=\exp(-2\pi |k|/H)$ which has common form with Unruh radiation and Hawking radiation. However, while for general $\alpha$-vacua, additional deviations from thermality should be included
\be
|0^{(\alpha)}_k\ra=\sqrt{1-\tanh^2rf^2}\sum_{n=0}^\infty\tanh^nrf^n|n_k^{(I)};n_{-k}^{(II)}\ra\label{2modealpha}
\ee
where
\be
f\equiv\frac{1+e^\alpha\tanh^{-1} r}{1+e^\alpha\tanh r}=\frac{1+e^{\alpha+\pi |k|/H}}{1+e^{\alpha-\pi |k|/H}}
\ee
As $\alpha\rightarrow-\infty$, these correction can be neglected and a pure thermal de Sitter radiation is left.
The one-particle excitation in $\alpha$-vacua can also be obtained as
\be
|1^{(\alpha)}_k\ra=\Big[1-\tanh^2rf^2\Big]\sum_{n=0}^\infty \tanh^nrf^n\sqrt{n+1}|n_k^{(I)};n_{-k}^{(II)}\ra
\label{onealpha}
\ee

\emph{Quantum entanglement degradation via vacuum ambiguity.}---The simplest way to illustrate the influence of vacuum ambiguity on quantum information task is to consider a maximally entangled state
\be
|\Psi\ra=\frac{1}{\sqrt2}(|0^{(\alpha)}_A\ra|0^{(\alpha)}_R\ra+|1^{(\alpha)}_A\ra|1^{(\alpha)}_R\ra)
\label{tripartite}
\ee
shared by two comoving observers Alice and Rob in coordinates (\ref{planar}) with respect to conformal time $\xi$. As Rob turns to be static, the corresponding states should be change into those of static coordinates. Since Rob is causally disconnected from region II, one obtains a mixed state after tracing over the states in region II,
\bea
\rho_{A,RI}
&=&\sum_{n=0}^{\infty}\frac{1}{2}\tanh^{2n}rf^{2n}(1-\tanh^2rf^2) \Big[\;|0n\rangle \langle 0n|\no\\
&&\hspace*{-40pt}+\sqrt{(n+1)(1-\tanh^2rf^2)}(|0n\rangle\langle 1n+1|+|1n+1\ra\langle
0n|)\no\\
&&\hspace*{-40pt}+ (n+1 )(1-\tanh^2rf^2)\times|1n+1\rangle \langle
1n+1|\;\Big]
\label{domAI}
\eea

To estimate the quantum correlations, we should calculate the negativity \cite{nega} which is entanglement monotone defined as the sum of negative eigenvalues of partial transposed density matrix. It follows that
\bea
N_{A,RI}&=&\sum_{n=0}^{+\infty}\frac{1}{4}\tanh^{2n}rf^{2n}(1-\tanh^2rf^2)\no\\
&&\hspace*{-30pt}\Bigg|\sqrt{[\tanh^2rf^2+n(\frac{\coth^2r}{f^2}-1)]^2+4(1-\tanh^2rf^2)}\no\\
&&\hspace*{-30pt}-\tanh^2rf^2+n\Big(\frac{1-\coth^2r}{f^2}\Big)\Bigg|
\label{negativityra}
\eea
which is a function of $\alpha$ and Hubble scale.

\begin{figure}[hbtp]
\includegraphics[width=.35\textwidth]{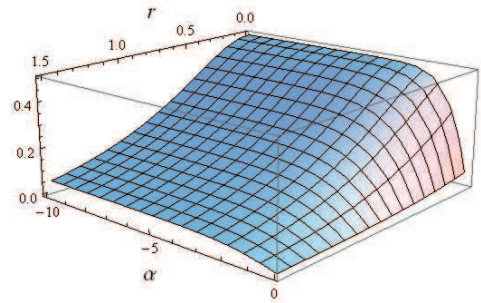}
\caption{The negativity of Alice-RobI system as the function of Hubble scale and $\alpha$. }
\label{negativity}
\end{figure}

Our first observation is that a degradation of quantum entanglement occur for static Rob (see Fig.\ref{negativity}) as we expected. This phenomena roots from the information lost via de Sitter radiation detected by inertial observer similar as Unruh effect in accelerated frame of flat space. For Euclidean state which means $\alpha=-\infty$, the spectrum is pure thermal. As $r\rightarrow0$ ($H\rightarrow0$), de Sitter space approaches flat with infinite large curvature radius, and the negativity (\ref{negativityra}) would approach 0.5 due to vanish radiation. On the other hand, in the limit of infinite curvature, the state has no longer distillable entanglement since negativity is exactly zero.

\begin{figure}[hbtp]
\includegraphics[width=.35\textwidth]{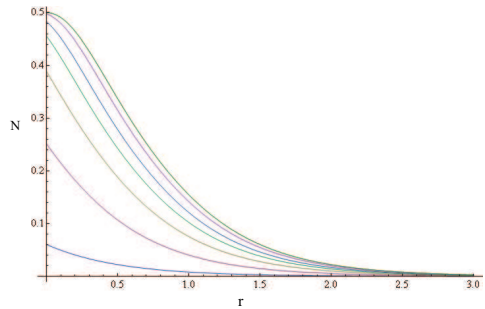}
\caption{The negativity as the superselection rule of vacua. For $\alpha\rightarrow-\infty$, negativity decay starts from 0.5, and is completely suppressed when $\alpha\rightarrow0^-$. }
\label{negativityalpha}
\end{figure}

The striking new phenomena is that the entanglement in de Sitter space can be directly related with vacua ambiguity (see Fig.\ref{negativityalpha}). For $\alpha\neq-\infty$, the negativity is suppressed compared with those observed by Rob when Euclidean vacuum be chosen. As $\alpha\rightarrow0^-$, the negativity becomes vanish and is independent with space curvature. It can be interpreted that one has a sudden vanishing of state's distillable quantum entanglement as Rob turns to be static. This result may seems pathologic at first glance since it indicates a sudden decrement of entanglement even for an inertial observer in the limit $H\rightarrow0$ where de Sitter space approaching flat. However it should be noted that in Minkowskian limit all $\alpha$-vacua with $\alpha\neq-\infty$ become squeezed states over Euclidean vacuum which matches the conformal vacuum of Minkowski space. Therefore, our results are consist with those similar arguments in flat space \cite{flat}. Moreover, since the quantum entanglement encodes the information of initial vacuum selection (i.e. a superselection rule of vacua), we expect that, in principle, one can determine in which right physical vacuum we stays by some quantum information tasks.

\begin{figure}[hbtp]
\includegraphics[width=.37\textwidth]{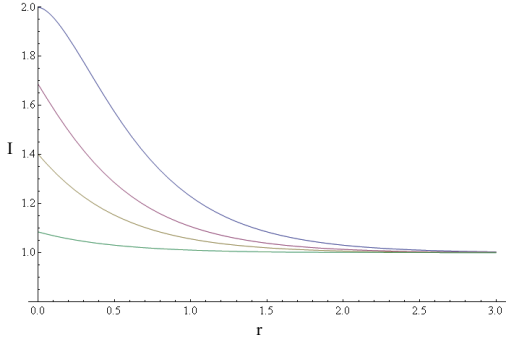}
\caption{Mutual information of Alice-RobI for various fixed $\alpha$.}
\label{mutualinf}
\end{figure}

We now estimate the mutual information which measures the total classical and quantum correlations in Alice-Rob system, $I_{A,RI}=S(\rho_A)+S(\rho_{RI})-S(\rho_{A,RI})$, where $S(\rho)=-\mbox{Tr}(\rho\log_2\rho)$ is the von Neumann entropy. A straightforward calculation gives
\bea
I_{A,RI}&=&1-\frac{1}{2}\log_{2}\left( \tanh ^{2}rf^2\right) -\frac{1}{2}(1-\tanh^2rf^2)\sum_{n=0}^{\infty}\no\\
&&\hspace*{-40pt}\tanh ^{2n}rf^{2n}\Big\{(1-n+n\frac{\coth^2r}{f^2})\log_2(1-n+n\frac{\coth^2r}{f^2})\no\\
&&\hspace*{-40pt}-[n+2-(n+1)\tanh^2rf^2]\no\\
&&\hspace*{-40pt}\times\log_2[n+2-(n+1)\tanh^2rf^2]\Big\}
\eea
which is a function of $\alpha$ and Hubble scale.
As the space curvature approaching infinite $H\rightarrow\infty$, $I\rightarrow1$, most quantum correlations have decayed. The mutual information should also encode different vacua selected by Alice-Rob initially. Therefore, we have a infinite set of mutual information evolution trajectories labeled by $\alpha$ continuously as illustrated in Fig.\ref{mutualinf}. Especially for $\alpha\neq-\infty$, a sudden decrement of $I$ due to spacetime curvature and vacua-selection occurs from Rob's view as he turning to be static. Since (\ref{tripartite}) is pure, one reads that $S(\rho_{A,RI})=S(\rho_{RII})$ and $S(\rho_{A,RII})=S(\rho_{RI})$. Therefore the combined mutual information, $I_{A,RI}+I_{A,RII}=2$, is conserved which suggests a correlation transfer between Alice-RobI and Alice-RobII systems and be independent with initial vacua selection.

\emph{Quantum discord.}---As learned from RQI in Minkowski space \cite{discord3}, there could be the residual quantum correlation measured by quantum discord of system exist even at infinite acceleration limit. Similar results should also be expected in de Sitter space as the distillable entanglement vanish. However, as we stated before, the behavior of quantum correlation is also influenced by initial vacuum selection. Therefore, it becomes necessary to specify the quantum discord evolution for distinct selected initial vacua.

To determine the quantum discord \cite{vedral1,vedral2,discord1} of our system which measures all quantum correlations left beside classical part in mutual information $D(A:RI)=I(A:RI)-C(A:RI)$, we first measure the subsystem of Alice of $\rho_{A,RI}$ by a complete set of projectors $\{\Pi_\pm=\frac{I_1\pm \vec{x}\cdot\vec{\sigma}}{2}\}$,
where $\vec{x}$ is parameterized as $(\sin\theta\cos\phi,\sin\theta\sin\phi,\cos\theta)$. The post-measurement density matrix becomes $\rho_{RI|\pm}=\mbox{Tr}(\Pi_j\rho_{A,RI}\Pi_j)/p_j$, and $p_j=\mbox{Tr}_{A,RI}(\Pi_j\rho_{A,RI}\Pi_j)=1/2$ for our case.

\begin{figure}[hbtp]
\includegraphics[width=.37\textwidth]{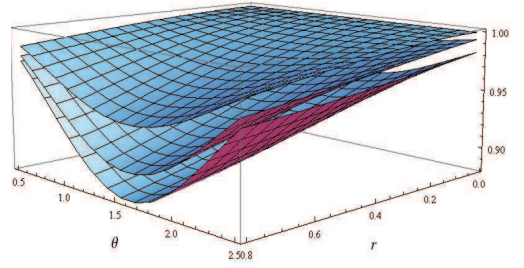}
\caption{Quantum discords of Alice-RobI as functions of $\theta$ and Hubble scale. Three hypersurfaces correspond to $\alpha=-20,-1,-0.5$ respectively.}
\label{discordAIar}
\end{figure}

\begin{figure}[hbtp]
\includegraphics[width=.37\textwidth]{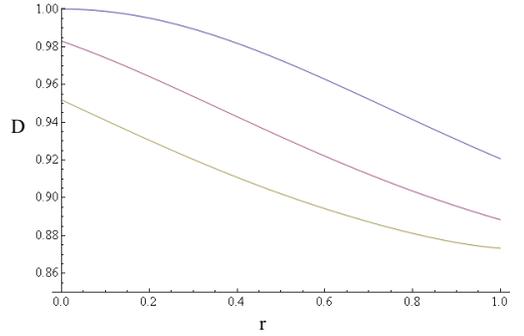}
\caption{Quantum discords of Alice-RobI as functions of Hubble scale. Three curves correspond to $\alpha=-20$, $\alpha=-1$ and $\alpha=-0.5$ respectively.}
\label{discordAIr}
\end{figure}

By minimize the conditional entropy $S_{\{\Pi_j\}}(RI|A)=\sum_jp_jS(\rho_{RI|\pm})$, one can rewrite the quantum discord as $\displaystyle D=S(\rho_A)-S(\rho_{ARI})+\min_{\{\Pi_j\}}S_{\{\Pi_j\}}(RI|A)$ and yield
\bea
D(\theta,r,\alpha)&=&1+\frac{1}{2}(1-\tanh^2rf^2)\sum_{n=0}^{\infty}\Big\{\tanh^{2n}rf^{2n}\no\\
&&\hspace*{-40pt}\times[1+(n+1)(1-\tanh^2rf^2)]\log_2\Big[\frac{1}{2}(1-\tanh^2rf^2)\no\\
&&\hspace*{-40pt}\times\tanh^{2n}rf^{2n}[1+(n+1)(1-\tanh^2rf^2)]\Big]\Big\}\no\\
&&\hspace*{-40pt}-\max\frac{1}{2}\sum_{i=\pm}\mbox{Tr}(\lambda_i\log_2\lambda_i)
\label{discordAI}
\eea
where $\lambda_i$ is in the eigenvalues spectrum of $\rho_{RI|\pm}$. As plotted in Fig.\ref{discordAIar} for three distinct initial vacua with $\alpha=-20,-1,-0.5$, the common minima appears at $\theta=\frac{\pi}{2}$. Therefore the quantum discord for Alice-RobI system is $D(\theta=\frac{\pi}{2},r,\alpha)$, depicted in Fig.\ref{discordAIr}. We read that the quantum discord decay as the $H\rightarrow\infty$ but reach a nonvanishing minimal value of quantum correlation, similar as the case for accelerator in Minkowski space. Moreover, one also reads distinct evolution trajectories of discord for $\alpha\neq-\infty$. In the limit of $\alpha\rightarrow0^-$, the quantum discord keep staying at its minima and becomes nonsensitive with spacetime curvature.


\emph{Experiment test by ion trap.}---
In realistic inflationary regime, $\la0^{(\alpha)}|\delta\phi^2|0^{(\alpha)}\ra$ leads to the observationally corroborated
primordial CMBR anisotropies \cite{COBE,transplanck}, restricts $\alpha$ as $e^\alpha\sim H/\Lambda$. The design of intelligent quantum information experiments to match these observations is the next challenging open problem. However, one can still simulate the quantum effects from vacuum ambiguity by analogue gravity experiments using ion trap.
In detector picture \cite{ion}, replacing the conformal time $\xi$ with the experimenter's clock time, the detector should evolve with respect to the simulated proper time $t$ which is equal to the cosmic time. To simulate vacuum ambiguity, the detector response function represents the probability for an ion excited by interaction with the field. For general $\alpha\neq-\infty$, the Wightman function is $G^+_{\alpha}(\xi,\xi')=N_\alpha[G^+_E(\xi,\xi')+e^{\alpha+\alpha^*}G^+_E(\xi',\xi)+e^\alpha G^+_E(-\xi,\xi')+e^{\alpha^*} G^+_E(\xi',-\xi)]$. The ion analogue of this function $\la\phi_m(\xi)\phi_m(\xi')\ra$ should be evaluated in some motional-state\cite{ionsqueeze}, since the $\alpha$-vacua can also be interpreted as squeezed states over Euclidean vacuum state.

\emph{Conclusions.}---It is shown that quantum correlation, in particular quantum entanglement, not only
 plays a key role in quantum information science but also can stimulate the interaction with other
 research areas such as condensed matter physics, for example \cite{cuinaturecomm}. In this Letter, we find that
quantum information may also be closely related with some fundamental problems in cosmology.
We have recognized the signature of vacuum ambiguity of de Sitter
space in various quantum correlations of scalar field modes. The significance of this result is indicated by
the possibility of study the vacuum ambiguity of cosmology from the view of quantum information methodology which is
very distinct with the traditional field-theoretical approach. This study may bridge the quantum information
science with quantum cosmology.



J. F has greatly benefitted from conversations with Bo Li. This work was supported by NSFC, ¡°973¡± program (2010CB922904).

\end{document}